\documentclass[12pt]{article}
\usepackage{cite}
\usepackage{amsfonts,latexsym}
\usepackage{epsfig}
\begin{document}
\newcommand{\de}{\delta}\newcommand{\ga}{\gamma}
\newcommand{\e}{\epsilon} \newcommand{\ot}{\otimes}
\newcommand{\be}{\begin{equation}} \newcommand{\ee}{\end{equation}}
\newcommand{\ba}{\begin{array}} \newcommand{\ea}{\end{array}}
\newcommand{\beq}{\begin{equation}}\newcommand{\eeq}{\end{equation}}
\newcommand{\tmod}{{\cal T}}\newcommand{\amod}{{\cal A}}
\newcommand{\bemod}{{\cal B}}\newcommand{\cmod}{{\cal C}}
\newcommand{\dmod}{{\cal D}}\newcommand{\hmod}{{\cal H}}
\newcommand{\s}{\scriptstyle}\newcommand{\tr}{{\rm tr}}
\newcommand{\einsop}{{\bf 1}}
\def\R{\overline{R}} \def\doa{\downarrow}
\def\dag{\dagger}
\def\ve{\epsilon}
\def\si{\sigma}
\def\ga{\gamma}
\def\no{\nonumber}
\def\le{\langle}
\def\re{\rangle}
\def\lt{\left}
\def\rt{\right}
\def\dwn{\downarrow} 
\def\up{\uparrow}
\def\dag{\dagger}
\def\nonum{\nonumber}
\newcommand{\reff}[1]{eq.~(\ref{#1})}

%%%%%%%%%%%%%%
\def\oR{R^*}
\def\upa{\uparrow}
\def\R{\overline{R}}
\def\doa{\downarrow}
\def\dag{\dagger}
\def\ve{\epsilon}
\def\si{\sigma}
\def\ga{\gamma}

% Shorthands for \begin{equation} and the like

\def\ba{\begin{array}}
\def\ea{\end{array}}
\def\no{\nonumber}
\def\le{\langle}
\def\re{\rangle}
\def\lt{\left}
\def\rt{\right}
\def\dwn{\downarrow}
\def\up{\uparrow}
\def\dag{\dagger}
%\def\nonum{\nonumber}

%%%%%%%%%%%%%%%%%
\title{Integrable impurity spin ladder systems}

\author{Arlei Prestes Tonel$^{1}$\footnote{E-mail: prestes@if.ufrgs.br},
Angela Foerster$^{1,2}$\footnote{E-mail: angela@if.ufrgs.br}, \\
Xi-Wen Guan$^1$\footnote{E-mail: guan@if.ufrgs.br}  and Jon Links$^{1,3}$ \footnote{E-mail: jrl@maths.uq.edu.au}
\vspace{1.0cm}\\
$^{1}$Instituto de F\'{\i}sica da UFRGS \\
Av. Bento Gon\c{c}alves 9500, Porto Alegre, RS - Brazil
\vspace{0.5cm}\\
$^{2}$Institut f\"ur Theoretische Physik \\
 Freie Universit\"at at Berlin, Arnimallee 14, 14195 Berlin, Germany
\vspace{0.5cm}\\
$^{3}$Department of Mathematics \\
The University of Queensland, Queensland, 4072,  Australia
}

\maketitle

\begin{abstract}
Two integrable spin ladder systems with different types of impurities
are proposed. The impurities are introduced in such a way that the
integrability of the models is not violated. The models are
solved exactly and the Bethe ansatz equations as well as the energy
eigenvalues are obtained.
We show for both models that a phase transition between gapped and
gapless
spin excitations occurs at a critical value of the rung
coupling $J$. In addition, the dependence of the impurities on
this
phase transition is determined explicitly. Remarkably, in one
of the models a decreasing of the spin gap with increasing  
impurity strength is
found.
\end{abstract}


\vfil\eject
%%%%%%%%%%%%%%%%%%%%%%%%%

The study of spin ladder systems continues to generate activity 
in order to understand the crossover from one-dimension to two-dimensions
in condensed matter physics.  
In fact, with the rapid progress presently being made in
nano-engineering,
several compounds have been realized experimentally with
a ladder structure (see for example \cite{dag,dago}). 
In addition, 
experiments using different techniques, such
as magnetic susceptibility measurements \cite{az} or 
nuclear magnetic ressonance \cite{ish},
report on the existence of a spin gap
in the spectrum of elementary excitations for even leg ladders.
The existence of such a spin gap is
crucial for superconductivity to occur under hole doping,
as verified experimentally in some of these compounds
and also predicted theoretically.

Initially, most of the theoretical results concerning ladder systems
were obtained
from studies of the standard Heisenberg
ladder. However, in contrast to its one-dimensional analogue,
this model can not be solved exactly.
Subsequently, in order to gain insight into the theory of
spin ladder systems, other models with generalized interaction
terms which guarantee integrability have been proposed 
\cite{wan,db,afw,bm1,frah,bm2,bdlm,dbm,links,kundu,ws,kat}.
Remarkably, such generalized models exhibit realistic physical
properties such as the existence of a spin gap \cite{wan} and the 
prediction of magnetization
plateaus at fractional values of the total magnetization \cite{db}.
However, although there is a considerable amount of work on
integrable spin ladder systems in the literature, very
few are concerned with the presence and influence of impurities.
(One exception however is \cite{ws}.)

The role of impurities in the context of spin chains and strongly
correlated electron systems is established to be an important 
aspect, particularly in 
low-dimensional cases.
When undertaking studies appealing to exact solutions afforded 
by integrable systems, it is possible to incorporate 
impurities into the system without violating
integrability. This can be achieved via two methods. 
In the first case, the impurities are given by changing
the representation of the underlying Lie algebraic structure
 at some lattice sites from the
fundamental representation to some other representation.
In this context, several versions of the Heisenberg and $t-J$ models,
for example, have been investigated  \cite{bed,beb1,flt,lf}.
Another possibility is to introduce
the impurities by way of inhomogeneities in the transfer matrix of the
system. This was explored, for instance in
\cite{andrei,sse,hpe,bares}.

Here we wish to adapt these two known methods in one-dimension
to incorporate impurities
to the quasi-bidimensional case for the purpose of studying 
impurities in spin ladder systems. In particular,
we will construct two integrable spin ladder models based on the
$su(4)$ algebra (Wang's model \cite{wan}) with impurities. The effect
of these impurities on the phase transition between the gapped
and gapless spin excitations of both models will be investigated.
We will show that in one of the cases the gap does not depend
explicitly on the impurity, while it does in the other case. 
Moreover, it turns out in the latter instance that the
spin gap decreases by increasing the impurity.

%%%%%%%%%%%%%%%%%%
Now, we begin by introducing the first model whose Hamiltonian reads
\be
H=\sum_{i=1}^{N}  \, h_{i,i+1} +
\frac{6}{\Lambda(\Lambda -8)}  \,
Q_{i,\bar{\i}} +
\frac{2}{\Lambda -8}  \,
h_{i,i+1}Q_{i,\bar{\i}}
-\frac{2}{\Lambda}Q_{i,\bar{\i}} h_{i,i+1}+
\frac{J}{2}\sum_{i=1}^{N}\left( \vec{\sigma}_{i}.\vec{\tau}_{i} - 1 
\right)
-\frac{J}{2} \left(\vec{\sigma}_{\bar{\i}}.\vec{\tau}_{\bar{\i}} - 1 \right),
\label{ham}
\ee
where $h_{i,i+1}= -\frac{1}{2} P_{i,i+1}$ with $P_{i,i+1}$ and
$Q_{i,\bar{\i}}$ given by

$$h_{i,i+1}=\frac{1}{4}\left(1+ \vec{\sigma}_{i}.\vec{\sigma}_{i+1} 
\right)
\left( 1+\vec{\tau}_{i}.\vec{\tau}_{i+1} \right)$$
$$Q_{i,\bar{\i}}=\frac{1}{4}[1+(\sigma_{i}^{x}\sigma_{\bar{\i}}^{x}-
\sigma_{i}^{y}\sigma_{\bar{\i}}^{y}+\sigma_{i}^{z}\sigma_{\bar{\i}}^{z})]
[ 1+(\tau_{i}^{x}\tau_{\bar{\i}}^{x}-\tau_{i}^{y}\tau_{\bar{\i}}^{y}+
\tau_{i}^{z}\tau_{\bar{\i}}^{z}) ].$$
Above $\vec{\sigma}_{i}$ and $\vec{\tau}_{i}$ are Pauli matrices acting
on the site $i$ (or the impurity site $\bar{\i}$) of the upper and lower 
legs of
the ladder, respectively, $J$  is the coupling constant
across the rungs (including the impurity rung) and $\Lambda$ is an arbitrary
parameter.
Throughout, $N$ is the number of rungs (equivalently, the length of the
ladder) and periodic boundary conditions are imposed. The picture below shows 
the 2-leg spin $1/2$ ladder in detail.

\[
\unitlength=0.50mm
\begin{picture}(20.,25.)
\put(-120,0){\line(1,0){240.}}
\put(-120,-50){\line(1,0){240.}}
\put(-100,-50){\line(0,1){50.}}
\put(-100.,0.){\makebox(0.,0.){$\bullet$}}
\put(-100.,-50.){\makebox(0.,0.){$\bullet$}}
\put(-110.,-25.){\makebox(0.,0.){$J$}}
\put(-100.,7.){\makebox(0.,0.){$i-1$}}
\put(-50,-50){\line(0,1){50.}}
\put(-50.,0.){\makebox(0.,0.){$\bullet$}}
\put(-50.,-50.){\makebox(0.,0.){$\bullet$}}
\put(-60.,-25.){\makebox(0.,0.){$J$}}
\put(-50.,7.){\makebox(0.,0.){$i$}}
\put(-0.5,-50){\line(0,1){50.}}
\put(0,-50){\line(0,1){50.}}
\put(0.,0.){\makebox(0.,0.){$\bullet$}}
\put(0.,-50.){\makebox(0.,0.){$\bullet$}}
\put(0.,7.){\makebox(0.,0.){$\bar{\i}$}}
\put(0.5,-50){\line(0,1){50.}}
\put(-10.,-25.){\makebox(0.,0.){$J$}}
\put(50,-50){\line(0,1){50.}}
\put(50.,0.){\makebox(0.,0.){$\bullet$}}
\put(50.,-50.){\makebox(0.,0.){$\bullet$}}
\put(50.,7.){\makebox(0.,0.){$i+1$}}
\put(40.,-25.){\makebox(0.,0.){$J$}}
\put(100,-50){\line(0,1){50.}}
\put(100.,0.){\makebox(0.,0.){$\bullet$}}
\put(100.,-50.){\makebox(0.,0.){$\bullet$}}
\put(100.,7.){\makebox(0.,0.){$i+2$}}
\put(90.,-25.){\makebox(0.,0.){$J$}}
\end{picture}
\vspace{0.50cm}
\]
\vspace{2.0cm}

Notice here that besides the usual Heisenberg interactions along the
legs and rungs we also have  Heisenberg  type interactions between the
lattice site $i$ and the impurity site $\bar{\i}$, as well
as biquadratic interactions (also involving the impurity site
$\bar{\i}$). In addition, there are three site
interactions, involving the sites $i$ and $i+1$ as well as the impurity
site $\bar{\i}$.
Finally, we mention that
the Hamiltonian for the ladder model based on the $su(4)$ symmetry
\cite{wan} can be recovered from eq. (\ref{ham}) by taking the
limit $\Lambda \rightarrow \infty$.

The energy eigenvalues of the Hamiltonian (\ref{ham}) are given by
\begin{equation}
\label{energy1}
E= 2J+(1-2J)N-4\sum_{i=1}^{M_{1}}\biggl(\frac{1}
{\lambda_{i}^{2}+1}-\frac{J}{2}
\biggr)
\end{equation}
where $\lambda_i$ are solutions to the Bethe ansatz equations(BAE)
below.
The BAE arise from the exact solution of the model through the
nested algebraic Bethe ansatz method and read

\begin{eqnarray}
\left(\frac{\lambda_{l}-i}
{\lambda_{l}+i}\right)^{N}&=&\prod_{l\neq
i}^{M_{1}}\frac{\lambda_{l}-\lambda_{i}-2i}
{\lambda_{l}-\lambda_{i}+2i}
\prod_{j=1}^{M_{2}}\frac{\lambda_{l}-\mu_{j}+i}
{\lambda_{l}-\mu_{j}-i} \nonumber \\
\prod_{j\neq l}^{M_{2}}\frac{\mu_{l}-\mu_{j}-2i}
{\mu_{l}-\mu_{j}+2i}&=&\prod_{i=1}^{M_{1}}\frac{\mu_{l}-\lambda_{i}-i}
{\mu_{l}-\lambda_{i}+i}
\prod_{k =1}^{M_{3}}\frac{\mu_{l}-\nu_{k}-i}
{\mu_{l}-\nu_{k}+i}
\label{baef} \\
\left(\frac{\nu_l -\Lambda +i}{\nu_l -\Lambda -i}\right)\prod_{k \neq
l}^{M_{3}}
\frac{\nu_{l}-\nu_{k}-2i}{\nu_{l}-\nu_{k}+2i} &=&
\prod_{j =1}^{M_{2}}\frac{\nu_{l}-\mu_{j}-i}
{\nu_{l}-\mu_{j}+i}
\nonumber
\end{eqnarray}
%%%%
\noindent The presence of the impurity can be detected
explicitly through the presence of the parameter $\Lambda$ in the first
term
of the last equation above.

Here the ground state is given by a product of  rung singlets  when 
$J >2$ and the energy is
$E_0= 2J +(1-2J)N$. This is in fact the reference state used in
the Bethe ansatz calculations and corresponds to the case
$M_1=M_2=M_3=0$ of the BAE (\ref{baef}). To describe an elementary
excitation, we choose $M_1=1 $ and $M_2=M_3=0$ in the BAE (\ref{baef})
which gives the minimal excited state energy,
$E_1=-4+4J+(1-2J)N.$

 The energy gap can easily be calculated and is found to be

\begin{equation}
\Delta=2\biggl(J-2\biggr)  .
\end{equation}
By solving $\Delta = 0$ for $J$ we find the critical value
$J^c=2$, indicating the critical
point at which the quantum phase transition from the
dimerized phase to the gapless phase occurs.
Notice that there is no effect of the impurity $\Lambda$
on the gap.
%%%%%%%

The integrability of this model (for $J=0$) can be shown by
the fact that there are
matrices $R$ and $R^{\ast}$ given by
\begin{equation}
\label{r}
{\footnotesize
R_{12}=
\begin{array}{c}
\unitlength=0.50mm
\begin{picture}(20.,25.)
\put(11.,-4.){\makebox(0.,0.){$\s 2 $}}
\put(24.,11.){\makebox(0.,0.){$\s 1 $}}
%\put(-2.,11.){\makebox(0.,0.){$\s \gamma $}}
%\put(11.,24.){\makebox(0.,0.){$\s 2 $}}
\put(0.,11.){\line(1,0){22.}}
\put(11.,21.){\line(0,-1){22.}}
\end{picture}
\end{array} =
\pmatrix{a&0&0&0&|& 0&0&0&0&|& 0&0&0&0&|& 0&0&0&0&\cr
               0&b&0&0&|& c&0&0&0&|& 0&0&0&0&|& 0&0&0&0&\cr
               0&0&b&0&|& 0&0&0&0&|& c&0&0&0&|& 0&0&0&0&\cr
               0&0&0&b&|& 0&0&0&0&|& 0&0&0&0&|& c&0&0&0&\cr
               -&-&-&-& & -&-&-&-& & -&-&-&-& & -&-&-&-&\cr
               0&c&0&0&|& b&0&0&0&|& 0&0&0&0&|& 0&0&0&0&\cr
               0&0&0&0&|& 0&a&0&0&|& 0&0&0&0&|& 0&0&0&0&\cr
               0&0&0&0&|& 0&0&b&0&|& 0&c&0&0&|& 0&0&0&0&\cr
               0&0&0&0&|& 0&0&0&b&|& 0&0&0&0&|& 0&c&0&0&\cr
               -&-&-&-& & -&-&-&-& & -&-&-&-& & -&-&-&-&\cr
               0&0&c&0&|& 0&0&0&0&|& b&0&0&0&|& 0&0&0&0&\cr
               0&0&0&0&|& 0&0&c&0&|& 0&b&0&0&|& 0&0&0&0&\cr
               0&0&0&0&|& 0&0&0&0&|& 0&0&a&0&|& 0&0&0&0&\cr
               0&0&0&0&|& 0&0&0&0&|& 0&0&0&b&|& 0&0&c&0&\cr
               -&-&-&-& & -&-&-&-& & -&-&-&-& & -&-&-&-&\cr
               0&0&0&c&|& 0&0&0&0&|& 0&0&0&0&|& b&0&0&0&\cr
               0&0&0&0&|& 0&0&0&c&|& 0&0&0&0&|& 0&b&0&0&\cr
               0&0&0&0&|& 0&0&0&0&|& 0&0&0&c&|& 0&0&b&0&\cr
               0&0&0&0&|& 0&0&0&0&|& 0&0&0&0&|& 0&0&0&a&\cr} } \, \, \,
,
\end{equation}
with
$$a=-x/2+1\, \, \, \, ,b=-x/2,\, \, \, \,  c=1,$$
\begin{equation}
\label{rest}
{\footnotesize
{R^*}_{12}=
\ba{c}
\unitlength=0.50mm
\begin{picture}(20.,25.)
\put(11.,-4.){\makebox(0.,0.){$\s 2 $}}
\put(24.,11.){\makebox(0.,0.){$\s 1 $}}
\put(0.,11.){\line(1,0){22.}}
\put(11.00,21.00){\circle{3.0}}
%\put(11.00,20.00){\circle{2.0}}
\put(11.00,19.00){\circle{3.0}}
%\put(11.00,18.00){\circle{2.0}}
\put(11.00,17.00){\circle{3.0}}
%\put(11.00,16.00){\circle{2.0}}
\put(11.00,15.00){\circle{3.0}}
%\put(11.00,14.00){\circle{2.0}}
\put(11.00,13.00){\circle{3.0}}
%\put(11.00,12.00){\circle{2.0}}
\put(11.00,11.00){\circle{3.0}}
%\put(11.00,10.00){\circle{2.0}}
\put(11.00,9.00){\circle{3.0}}
%\put(11.00,8.00){\circle{2.0}}
\put(11.00,7.00){\circle{3.0}}
%\put(11.00,6.00){\circle{2.0}}
\put(11.00,5.00){\circle{3.0}}
%\put(11.00,4.00){\circle{2.0}}
\put(11.00,3.00){\circle{3.0}}
%\put(11.00,2.00){\circle{2.0}}
\put(11.00,1.00){\circle{3.0}}
%\put(11.00,0.00){\circle{2.0}}

\end{picture}
\ea
=\pmatrix{e&0&0&0&|& 0&d&0&0&|& 0&0&d&0&|& 0&0&0&d&\cr
               0&c&0&0&|& 0&0&0&0&|& 0&0&0&0&|& 0&0&0&0&\cr
               0&0&c&0&|& 0&0&0&0&|& 0&0&0&0&|& 0&0&0&0&\cr
               0&0&0&c&|& 0&0&0&0&|& 0&0&0&0&|& 0&0&0&0&\cr
               -&-&-&-& & -&-&-&-& & -&-&-&-& & -&-&-&-&\cr
               0&0&0&0&|& c&0&0&0&|& 0&0&0&0&|& 0&0&0&0&\cr
               d&0&0&0&|& 0&e&0&0&|& 0&0&d&0&|& 0&0&0&d&\cr
               0&0&0&0&|& 0&0&c&0&|& 0&0&0&0&|& 0&0&0&0&\cr
               0&0&0&0&|& 0&0&0&c&|& 0&0&0&0&|& 0&0&0&0&\cr
               -&-&-&-& & -&-&-&-& & -&-&-&-& & -&-&-&-&\cr
               0&0&0&0&|& 0&0&0&0&|& c&0&0&0&|& 0&0&0&0&\cr
               0&0&0&0&|& 0&0&0&0&|& 0&c&0&0&|& 0&0&0&0&\cr
               d&0&0&0&|& 0&d&0&0&|& 0&0&e&0&|& 0&0&0&d&\cr
               0&0&0&0&|& 0&0&0&0&|& 0&0&0&c&|& 0&0&0&0&\cr
               -&-&-&-& & -&-&-&-& & -&-&-&-& & -&-&-&-&\cr
               0&0&0&0&|& 0&0&0&0&|& 0&0&0&0&|& c&0&0&0&\cr
               0&0&0&0&|& 0&0&0&0&|& 0&0&0&0&|& 0&c&0&0&\cr
               0&0&0&0&|& 0&0&0&0&|& 0&0&0&0&|& 0&0&c&0&\cr
               d&0&0&0&|& 0&d&0&0&|& 0&0&d&0&|& 0&0&0&e&\cr} } \, \, \,
,
\end{equation}
with
$$c=1\, \, \, \, ,d=2/x,\, \, \, \,  e=1+\frac{2}{x},$$
obeying the Yang-Baxter algebra
\begin{equation}
R_{12}(x-y)R_{13}(x)R_{23}(y)=
R_{23}(y)R_{13}(x)R_{12}(x-y).
\end{equation}
\begin{equation}
R_{12}(x-y)R_{13}^{\ast}(x)R_{23}^{\ast}(y)=
R_{23}^{\ast}(y)R_{13}^{\ast}(x)R_{12}(x-y).
\end{equation}
Above $R(x)$ is the fundamental $R$-matrix associated to the $su(4)$ algebra
while $R^{\ast}(x)$ is the solution which acts in the tensor product of 
the fundamental representation and its dual.
By applying the standard procedure of the Quantum Inverse Scattering
Method (QISM), the global Hamiltonian is obtained from
\begin{equation}
H=\frac{d}{dx}\ln (\tau (x,\Lambda))\vert _{x=0}
\label{haf}
\end{equation}
where $\tau (x,\Lambda)$ is the transfer matrix
\begin{equation}
\tau (x,\Lambda)= {\rm tr}_0 (R_{0,1}(x)\ldots
R_{0,i}(x)R_{0,\bar{\i}}^{\ast}(x-\Lambda)R_{0,i+1}(x)\ldots R_{0,N}(x))
\end{equation}
and ${\rm tr}_0$ denotes the trace over the auxiliary space, labelled by $0$.

\[
\unitlength=0.50mm
\begin{picture}(20.,25.)
\put(-55.,0.){\line(1,0){110.}}
\put(-135.,0.){\line(1,0){35.}}
\put(142.,0.){\makebox(0.,0.){$0$}}
\put(-130.,-25.){\line(0,1){50.}}
\put(-105.,-25.){\line(0,1){50.}}
\put(-105.,-32.){\makebox(0.,0.){$2$}}
\put(-135.,-32.){\makebox(0.,0.){$1$}}
\put(-75.,0.){\makebox(0.,0.){$\ldots $}}
\put(-85.,0.){\makebox(0.,0.){$\ldots $}}
\put(95.,0.){\line(1,0){40.}}
\put(100.,-25.){\line(0,1){50.}}
\put(100.,-32.){\makebox(0.,0.){$N-1$}}
\put(130.,-25.){\line(0,1){50.}}
\put(130.,-32.){\makebox(0.,0.){$N$}}
\put(65.,0.){\makebox(0.,0.){$\ldots $}}
\put(75.,0.){\makebox(0.,0.){$\ldots $}}
\put(85.,0.){\makebox(0.,0.){$\ldots $}}
\put(-65.,0.){\makebox(0.,0.){$\ldots $}}
\put(-30.,-25.){\line(0,1){50.}}
\put(-30.,-32.){\makebox(0.,0.){$i$}}
%\put(-10.,-25.){\line(1,2){26.93}}
\put(-10.,-25.){\circle{3.0}}
%\put(0.,0.){\circle{3.0}}
\put(9.20,23.){\circle{3.0}}
\put(8.40,21.){\circle{3.0}}
\put(10.,25.){\circle{3.0}}
\put(7.60,19.){\circle{3.0}}
\put(6.80,17.){\circle{3.0}}
\put(6.0,15.){\circle{3.0}}
\put(5.20,13.){\circle{3.0}}
\put(4.40,11.){\circle{3.0}}
\put(3.60,9.){\circle{3.0}}
\put(2.80,7.){\circle{3.0}}
\put(2.,5.){\circle{3.0}}
\put(1.20,3.){\circle{3.0}}
\put(0.40,1.){\circle{3.0}}
\put(-0.40,-1.){\circle{3.0}}
\put(-1.20,-3.){\circle{3.0}}
\put(-2.0,-5.){\circle{3.0}}
\put(-2.80,-7.){\circle{3.0}}
\put(-3.60,-9.){\circle{3.0}}
\put(-4.40,-11.){\circle{3.0}}
\put(-5.20,-13.){\circle{3.0}}
\put(-6.0,-15.){\circle{3.0}}
\put(-6.80,-17.){\circle{3.0}}
\put(-7.60,-19.){\circle{3.0}}
\put(-8.40,-21.){\circle{3.0}}
\put(-9.20,-23.){\circle{3.0}}
\put(-10.0,-25.){\circle{3.0}}
\put(-10.,-32.){\makebox(0.,0.){$\bar{\i}$}}
\put(30.,-25.){\line(0,1){50.}}
\put(30.,-32.){\makebox(0.,0.){$i+1$}}
\end{picture}
\]
\vspace{2.0cm}

Notice here that the impurity is incorporated into the system through
the inclusion of the operator $R^{\ast}(x-\Lambda)$ in the transfer matrix
and
its effect on the spectrum can be detected by the presence of the
extra parameter $\Lambda$.
%%%%%%%%%%%

In this way, the Hamiltonian (\ref{ham}) can be mapped to the following 
Hamiltonian, which can be derived from the $R$ and $R^{\ast} $ 
matrix (eqs.(\ref{r},\ref{rest})) obeying the Yang-Baxter 
algebra for $J=0$, while for $J\ne 0$ the rung interaction takes the 
form of a chemical potential term, commuting with the Hamiltonian.

\be
\bar{H}=\sum_{i=1}^{N}  \, \bar{h}_{i,i+1} +
\frac{6}{\Lambda(\Lambda -8)}  \,
\bar{Q}_{i,\bar{\i}} +
\frac{2}{\Lambda -8}  \,
\bar{h}_{i,i+1}\bar{Q}_{i,\bar{\i}}
-\frac{2}{\Lambda}\bar{Q}_{i,\bar{\i}} \bar{h}_{i,i+1}-
2J\sum_{i=1}^{N} X_i^{00} + 2J X_{\bar{\i}}^{00},
\label{ham1}
\ee

where

$$\bar{h}_{i,i+1}=\sum_{\alpha , \beta =0}^{3} X_{i}^{\alpha \beta} X_{i+1}^{\beta \alpha}$$
$$\bar{Q}_{i,\bar{\i}}=\sum_{\alpha ,\beta =0}^{3} X_{i}^{\alpha \beta} X_{\bar{\i}}^{\alpha \beta}.$$
In the above, $ X^{\alpha \beta}_{i} =
|\alpha_{i}\re\le\beta_{i}|$ are the Hubbard operators with $
|\alpha_{i}\re$ being the orthogonalised eigenstates of the local

operator $(\vec{\sigma_{i}}{\bf .} \vec{\tau_{i}})$.
% The 
%Hamiltonians (\ref{ham1}) and (\ref{ham}) can be mapped one with other
%through a similarity  transformation.
%\begin{eqnarray}
%|\up,\up\re & \rightarrow &  |\up,\up\re, \nonum \\
%|\up,\dwn\re &\rightarrow & 1/\sqrt{2} ( |\up,\dwn\re - |\dwn,\up\re), \nonum %\\
%|\dwn,\up \re &\rightarrow & 1/\sqrt{2} ( |\up,\dwn\re + |\dwn,\up\re), \nonum% \\
%|\dwn,\dwn\re . &\rightarrow &|\dwn,\dwn\re. \label{basistransf}
%\end{eqnarray}
%%%%%%%%%%%%%%%%
%For $J=J^{\prime}=0$ the Hamiltoniam above (\ref{haf}) can be mapped
%into the
%Hamiltonian (\ref{ham}),
%while for $J, J^{\prime} \neq 0$ the same mapping takes the rung
%interaction terms
%to a  chemical potential form. Consequently, the model can be solved exactly, %as
%shown before.

%%%%%%%%%%%%%%%%%%%%%%%%%%%

Let us now introduce another spin ladder model with an impurity,
whose Hamiltonian reads
\be
H=\sum_{k=1}^{i-1}  \, h_{k,k+1} +
\sum_{k=i+1}^{N}  \, h_{k,k+1}
+ H_{imp} +
\frac{J}{2}\sum_{k=1}^{N}\left( \vec{\sigma}_{k}.\vec{\tau}_{k}-1
\right) \,
+ \frac{J}{2}\left(\vec{\sigma}_{i}.\vec{\tau}_{\bar{\i}}-1\right)
\label{ha}
\ee
where $H_{imp}$ represents the interaction of the chain with an impurity
localized in the $\bar{\i}$ position and is given by

\begin{eqnarray}
&H_{imp}&=\frac{4}{4-\Lambda^2}\biggl\{ P_{i,\bar{\i}}+
P_{\bar{\i},i+1}-
\frac{\Lambda^2}{4}P_{i,i+1}\biggr\} \quad \nonumber \\ & &+
\frac{2\Lambda}{4-\Lambda^2}\biggl\{ P_{i,i+1}
P_{i,\bar{\i}}-P_{i,\bar{\i}}P_{i,i+1}\biggr\}-
\frac{2\Lambda}{4-\Lambda^2}I_{i,\bar{\i}}\quad \nonumber \\ & &
\end{eqnarray}
Above, $h_{k,k+1}=P_{k,k+1}$ represents the permutation operator
between the lattice  sites $k$ and $k+1$
and has the form
$$P_{k,k+1}=\frac{1}{4}\left(1+ \vec{\sigma}_{k}.\vec{\sigma}_{k+1}
\right)\left( 1+\vec{\tau}_{k}.\vec{\tau}_{k+1} \right)$$
\noindent This term contains Heisenberg interactions along the
legs and rung, as well as biquadratic interactions, whose
physical importance has been pointed out in \cite{biq}.
Here we also have three site
interactions involving the lattice sites $i,i+1$ as well as
the impurity site $\bar{\i}$.

%%%%%%%%%%%%%%%%%%%%%%
Finally, we mention that Wang's ladder system \cite{wan}
with length $N+1$ can be recovered by taking
the limit $\Lambda = 0$.

The energy eigenvalues of the Hamiltonian (\ref{ha}) are
given by
\begin{equation}
\label{energy}
E=-2J+(1-2J)N
+\frac{2}{2+\Lambda}-4\sum_{i=1}^{M_{1}}\biggl(\frac{1}{\lambda_{i}^{2}
+1}-\frac{J}{2}\biggr)
\end{equation}
where $\lambda_{i}$ are solutions to the Bethe ansatz equations below
\begin{eqnarray}
\left(\frac{\lambda_{l}-i}
{\lambda_{l}+i}\right)^{N}\left(\frac{\lambda_{l}-i(1-\Lambda )}
{\lambda_{l}+i(1+\Lambda )}\right)&=&\prod_{l\neq
i}^{M_{1}}\frac{\lambda_{l}-\lambda_{i}-2i}
{\lambda_{l}-\lambda_{i}+2i}
\prod_{j=1}^{M_{2}}\frac{\lambda_{l}-\mu_{j}+i}
{\lambda_{l}-\mu_{j}-i} \nonumber \\
\prod_{j\neq l}^{M_{2}}\frac{\mu_{l}-\mu_{j}-2i}
{\mu_{l}-\mu_{j}+2i}&=&\prod_{i=1}^{M_{1}}\frac{\mu_{l}-\lambda_{i}-i}
{\mu_{l}-\lambda_{i}+i}
\prod_{k =1}^{M_{3}}\frac{\mu_{l}-\nu_{k}-i}
{\mu_{l}-\nu_{k}+i}
\label{bae} \\
\prod_{k \neq l}^{M_{3}}
\frac{\nu_{l}-\nu_{k}-2i}{\nu_{l}-\nu_{k}+2i} &=&
\prod_{j =1}^{M_{2}}\frac{\nu_{l}-\mu_{j}-i}
{\nu_{l}-\mu_{j}+i}
\nonumber
\end{eqnarray}
\noindent For $J > \frac{2}{\lambda^2+1}$ 
(see eqs. (\ref{lamb}, \ref{gap}) below)
the ground state is given by the product of rung singlets and the
energy is $-2J+\frac{2}{2+\Lambda}+(1-2J)N 
$.
 This is again the reference state used in the
Bethe ansatz calculation and corresponds to the case
$M_1=M_2=M_3=0$ for the BAE (\ref{bae}). To describe an
elementary  excitation, we take $M_1=1$ and $M_2=M_3=0$
in the BAE (\ref{bae}), which leads to the following solution
for the variable $\lambda$ \footnote{strictly, the lattice length
is assumed odd, such that in the limit where there is no impurity
$(\Lambda =  0)$ the correct gap expression discussed in \cite{wan}
can be recovered}

\begin{equation}
\label{lamb}
\lambda=\frac{\Lambda}{4}-\frac{1}{24}\biggl\{
(1+i\sqrt{3})K^{1/3}-\frac{1}{8}
(1-i\sqrt{3})(3\Lambda^2+16)K^{-1/3}\biggr\}
\end{equation}
with $K$ given by
$$K=27\Lambda^3+12i\sqrt{81\Lambda^4 +432\Lambda^2 +768}.$$
Notice that the solution of the BAE $\lambda$ depends explicitly on the 
impurity $\Lambda$.
This relation, although complicated, is illustrated in Fig. 1 below.

\begin{figure}
\begin{center}
\epsfig{file=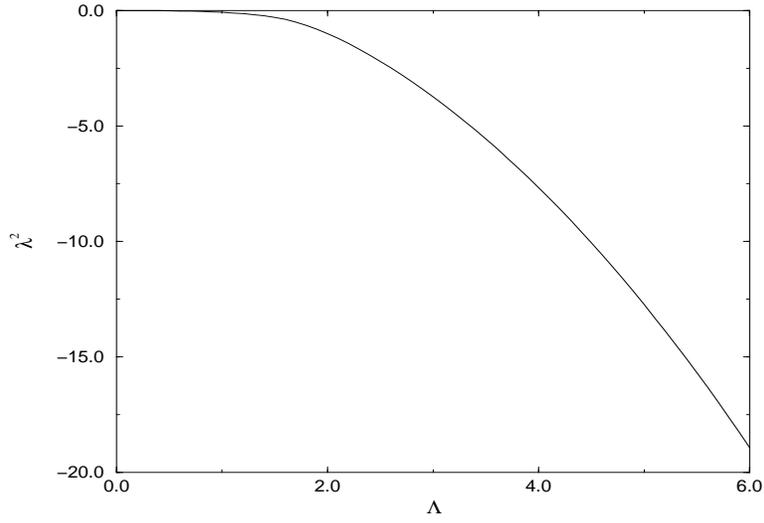,width=7cm,height=10cm,angle=-90}
\caption{This graph shows how the solution of the BAE ( $\lambda^2$)
 depends on the impurity  $\Lambda $.}
\end{center}          
\end{figure}

The energy gap can be calculated using the Bethe ansatz solution and has the 
form
\begin{equation}
\Delta=2\biggl(J-\frac{2}{\lambda^2+1}\biggr)  .
\label{gap}
\end{equation}
Here, in contrast to the previous case,
the impurity does affect the gap.
By solving $\Delta = 0$ for $J$ we find the critical value
$J^c=\frac{2}{\lambda^2+1}$, indicating the critical
line at which the quantum phase transition from the
dimerized phase to the gapless phase occurs. The phase transition line
is shown in Fig. 2.

\begin{figure}
\begin{center}
\epsfig{file=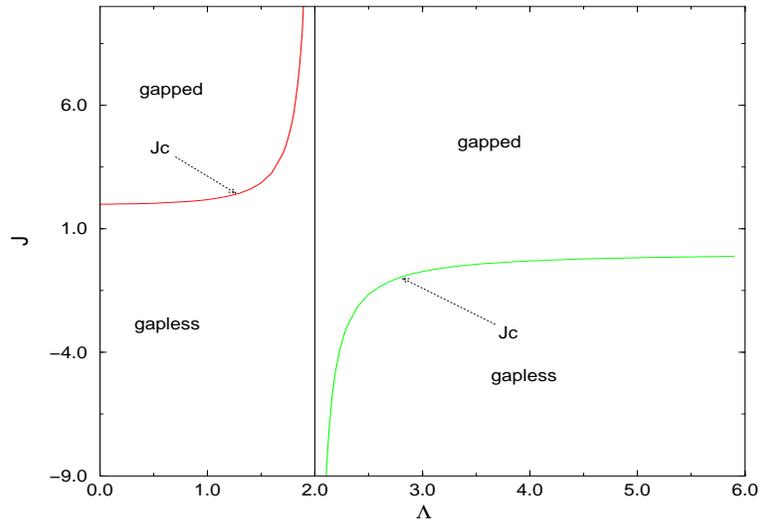,width=7cm,height=10cm,angle=-90}\\          
\caption{ Rung coupling $J$ versus impurity  $\Lambda$.
This graph represents the phase diagram. The curve 
($J^c=2/(\lambda^2+1) $) divides the gapped
and gapless phases.}
\end{center}
\end{figure}

Notice that by increasing the impurity $\Lambda$,
the critical value $J^c$ also increases.
% We can see here that
% all values $\Lambda > 2 $ correspond to negative rung coupling $J$.
% Since we are only interested in antiferromagnetic coupling,
% from now on we will restrict the discussion in the region
% $\Lambdda ??? [0,2) $ only.
A further analysis of both graphs together with the gap expression
(\ref{gap}) reveals that there is a reduction of the gap by
increasing $\Lambda$. This result can be easily confirmed by inspection
of  
Fig. 3.
%%%%%%%%%
\begin{figure}
\begin{center}
\epsfig{file=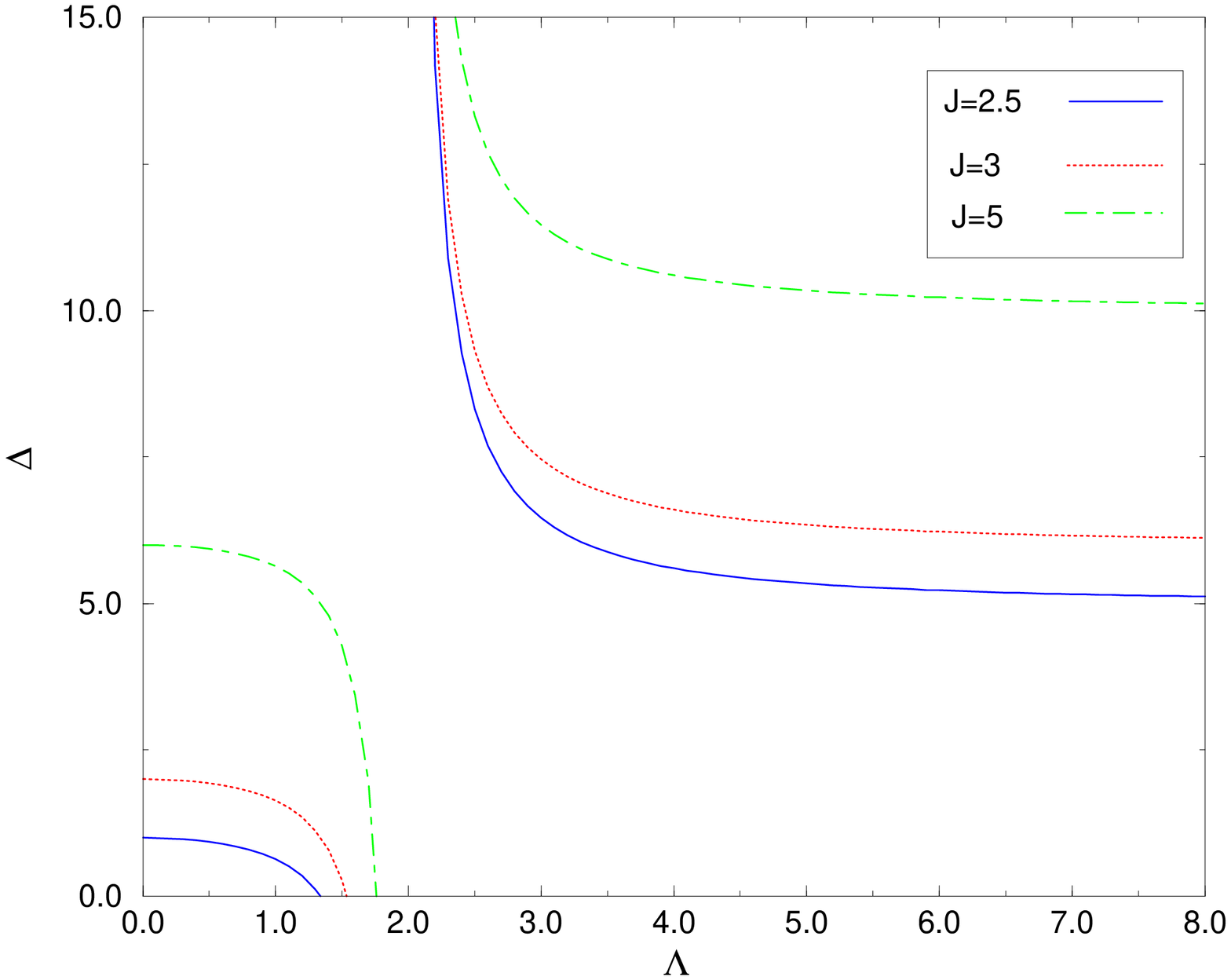,width=7cm,height=10cm,angle=-90}\\          
\caption{ This graph shows how the spin gap $\Delta$ depends on the 
impurity $\Lambda$ for different values of the rung coupling $J$. }
\end{center}
\end{figure}
%%%%%%%%%%%%
%\begin{figure}
%\begin{tabular}{cc}
%            &             \\
%\epsfig{file=gap1.eps,width=7cm,height=6cm,angle=-90}&            
%\epsfig{file=gap2.eps,width=7cm,height=6cm,angle=-90}\\
%  
%\end{tabular}
%\caption{The left graph 
%represents the dependence of the gap with $0<\Lambda <2$.
%The right graphic represents the dependence of the gap with $\Lambda >2$. In both 
%cases there is a decreasing of the gap with  the increasing of value of the impurity.}
%\end{figure}
%%%%%%%%%%%%%%%%%%%%
%%%%%%%%%

The exact solvability of this model, as for the previous case, can be 
shown by
the fact that for  $J=0$ 
there is a $R$-matrix given by
equation (\ref{r})  obeying Yang-Baxter algebra
\begin{equation}
R_{12}(x-y)R_{13}(x)R_{23}(y)=R_{23}(y)R_{13}(x)R_{12}(x-y).
\end{equation}

By the standard procedure the global Hamiltonian reads

\begin{equation}
H=\frac{d}{dx}\ln(\tau (x,\Lambda))\vert _{x=0}
\label{ha1}
\end{equation}
where $\tau (x,\Lambda)$ is the transfer matrix

\begin{equation}
\tau (x,\Lambda)= {\rm tr}_0 (R_{0,1}(x)\ldots 
R_{0,i}(x)R_{0,\bar{\i}}(x-\Lambda)R_{0,i+1}(x)\ldots R_{0,N}(x)
\end{equation}

\[
\unitlength=0.50mm
\begin{picture}(20.,25.)
\put(-55.,0.){\line(1,0){110.}}
\put(-135.,0.){\line(1,0){35.}}
\put(142.,0.){\makebox(0.,0.){$0$}}
\put(-130.,-25.){\line(0,1){50.}}
\put(-105.,-25.){\line(0,1){50.}}
\put(-105.,-32.){\makebox(0.,0.){$2$}}
\put(-135.,-32.){\makebox(0.,0.){$1$}}
\put(-75.,0.){\makebox(0.,0.){$\ldots $}}
\put(-85.,0.){\makebox(0.,0.){$\ldots $}}
\put(95.,0.){\line(1,0){40.}}
\put(100.,-25.){\line(0,1){50.}}
\put(100.,-32.){\makebox(0.,0.){$N-1$}}
\put(130.,-25.){\line(0,1){50.}}
\put(130.,-32.){\makebox(0.,0.){$N$}}
\put(65.,0.){\makebox(0.,0.){$\ldots $}}
\put(75.,0.){\makebox(0.,0.){$\ldots $}}
\put(85.,0.){\makebox(0.,0.){$\ldots $}}
\put(-65.,0.){\makebox(0.,0.){$\ldots $}}
\put(-30.,-25.){\line(0,1){50.}}
\put(-30.,-32.){\makebox(0.,0.){$i$}}
\put(-10.,-25.){\line(1,2){24.}}
\put(-10.,-32.){\makebox(0.,0.){$\bar{\i}$}}
\put(30.,-25.){\line(0,1){50.}}
\put(30.,-32.){\makebox(0.,0.){$i+1$}}
\end{picture}
\]
\vspace{2.0cm}

The shift that appears in $R_{0,\bar{\i}}(x-\Lambda)$ represents the impurity.

%%%%%%%
 The Hamiltonian (\ref{ha}) can be mapped, as in the previous case to the Hamiltonian  that is obtained by the  standard procedure from  $R$-matrix (\ref{r}) that obeys the Yang-Baxter algebra for $J=0$, while for $J\ne 0$ the rung interaction take the form of a chemical-potencial term. 
%%%%%%%%%%%%%%%

 To summarize, we have introduced two spin ladder models with an 
impurity $\Lambda$ . In the limit of vanishing impurity both 
models reduce to that introduced by Wang \cite{wan} 
 based on the $su(4)$ symmetry. The Bethe ansatz equations as well as
 the energy expression for the models were presented. We have shown that in 
one of the
 cases the impurity affects the gap of the system non-trivially, 
while in the other case there is no dependence. 

\vspace{2.0cm}

\centerline{{\bf Acknowledgements}}
~\\
APT  thanks CNPq-Conselho Nacional de Desenvolvimento
Cient\'{\i}fico e Tecnol\'ogico for financial support. AF thanks DAAD - Deutscher Akademischer Austauschdienst and FAPERGS - Funda\c{c}\~ao de Amparo a Pesquisa do Estado do Rio Grande do Sul for financial support. XWG thanks FAPERGS for financial support. JL thanks the ARC - Australian Research Council  for financial support. AF thanks the Institut f\"ur Theoretische Physik der Freie Universit\"at Berlin for their kind hospitality.

%%%%%%%%%%%%%%%%%%%%%%

\end{document}